\pdfoutput=1
\documentclass{llncs}
\usepackage{llncsdoc}
\usepackage{verbatim}
\usepackage{graphicx}
\usepackage{subfigure}
\usepackage{graphics}
\usepackage{multirow}
\usepackage{capt-of}
\usepackage{caption}

\begin{document}

\title{PicHunt: Social Media Image Retrieval for Improved Law Enforcement}
\author{Sonal Goel, Niharika Sachdeva, Ponnurangam Kumaraguru, \\A V Subramanyam \and Divam Gupta}
\institute{Indraprastha Institute of Information Technology, Delhi, India\\
$\left\{sonal1426,niharikas,pk,subramanyam,divam14038\right\}$@iiitd.ac.in}
\maketitle
\vspace{-0.2cm}
\begin{abstract}
First responders are increasingly using social media to identify and reduce crime for well-being and safety of the society. Images shared on social media hurting religious, political, communal and other sentiments of people, often instigate violence and create law \& order situations in society. This results in the need for first responders to inspect the spread of such images and users propagating them on social media.
In this paper, we present a comparison between different hand-crafted features and a Convolutional Neural Network (CNN) model to retrieve similar images, which outperforms state-of-art hand-crafted features. We propose an Open-Source-Intelligent (OSINT) real-time image search system, robust to retrieve modified images that allows first responders to analyze the current spread of images, sentiments floating and details of users propagating such content. The system also aids officials to save time of manually analyzing the content by reducing the search space on an average by 67\%. 

\end{abstract}

\vspace{-0.9cm}
\section{Introduction}
\vspace{-0.3cm}
First responders across the globe are increasingly using Online Social Networks (OSN) for maintaining safety and law \& order situations in society. Prior work shows the role of social media to aid first responders like police, for instance; police can use OSN to obtain actionable information like location, place, and evidence of the crime \cite{sachdeva2015social}. Police have realized the effectiveness of OSN in various activities such as investigation, crime identification, intelligence development, and community policing \cite{crump2011police,gerber2014predicting,sachdeva2015online}. However, in order to accomplish these goals on OSN, they often needs to identify tweets and images causing safety issues \cite{gerber2014predicting,wang2012automatic}. \par
Researchers have explored the utility of social media platforms like Twitter to identify text and network of people leading to law enforcement help \cite{crump2011police,wang2012automatic}, similarly images on social media sites also often yield information of interest to investigators \cite{govtech}.  A study shows that people don't engage equally with every tweet, Twitter content of over 2 million tweets by thousands of verified users over the course of a month was analyzed, and it showed that tweet with an image present can increase user engagement by 35\% \cite{Rogers_2014}. It is said that an image is worth a thousand words. People with different backgrounds can easily understand the main content of an image thus, increasingly becoming preferred media to reach and effect masses. 
\par
Often images shared on social media have the potential to hurt religious, political, communal, caste and other sentiments of a certain section of society. Such images intimidate people, instigate anger among them which further leads to law and order situations and security critical scenarios in the society. For example, in April 2016, a journalist tweeted a morphed image of an Indian politician touching feet of a foreign country's king, the image invited anger and backlash on all social media networks and the political party filed a complaint against the journalist for misleading public \cite{modi}. Another example, in June 2014, obscene pictures of Indian warrior-king Chhatrapati Shivaji and a late political party chief were posted on Facebook leading to riots in Maharastra, India. People went on rampage damaging public and private vehicles, pelting stones leading to severe injuries and even beating a person death \cite{shivaji}.   

\par

With this arises the need for first responders to understand, patrol, prevent the spread of such images for maintaining law and order in the society. 
However, police personnel has limited exposure to technology \cite{ecb_2014} and this limitation makes it difficult to adopt findings of OSN use by police to facilitate policing needs \cite{sachdeva2015online}. Moreover, researchers have focussed more on textual content to aid the first responders but analyzing multimedia content like images and videos on OSN is still largely unexplored \cite{kharroub2015social}. Also, the tools necessary to retrieve, filter, integrate and intelligently present relevant images and their information for better safety during security critical scenarios need to be leveraged \cite{cui2014social,hoi2011sire}. These initiatives motivated us to propose a real-time image-search system which can serve the first responders by bridging the gap between research in technologies and solving real-world problems to improve security during such critical scenarios. But the modifications done on images like cropping, scaling, stitching image, wrapping around text, contrast enhancement and changing colors are one of the biggest challenges faced in image analysis. These challenges often create barriers to directly access these images and utilize information like image spread, etc., which makes image retrieval a complex and daunting process. This paper lies in the intersection of crisis informatics and social media image utility in the understudied \& novel context of law enforcement. The main contributions of this work are:
\vspace{-0.23cm}
\begin{enumerate}
  \item We develop a real-time image search system, which is robust to detect similar images which are cropped, scaled, blurred, stitched with other images, wrapped around with text, brightened, or modified using other similar image processing techniques. 
  \item The system aims at data management for first responders, helping them reduce search space for images.
\item We analyze different techniques for retrieving similar images and experimentally show that ORB (Oriented Fast and Rotated Brief) in combination with RANSAC is the state-of-art technique in hand-crafted image features for identifying similar images.
  \item We propose a supervised deep CNN model that outperforms state-of-art hand-crafted techniques to retrieve similar images.
  \item We created a new human-annotated dataset of images from incidents that created law and order situations in the society.\footnote[1]{http://precog.iiitd.edu.in/resources.html}
\end{enumerate}

\vspace{-0.65cm}
\section{Problem Statement}
\vspace{-0.45cm}
We now formally define the problem definition and notations. Given an image $I_{A}$ and set of keywords $K$ by a user, find a set of similar images $I_{B} = \{I_{b1}, I_{b2},\dots I_{bn}\}$ using a comparison function $C$, from a search set of images $I_{C} = \{ I_{c1}, I_{c2},\dots I_{cm}\}$ formed using a search function $ S $ on a social network $S_{N}$, where $n \leq m$:
 \begin{center} $I_{C} = S(S_{N},K)$ and $ I_{B} = C(I_{A}, I_{C})$\end{center}

 The search function $S$ takes a set of keywords, and a search space $S_{N}$ as input and gives a search set of images $I_{C}$. The comparison function takes the given image  $I_{A}$ and $I_{C}$ as input and returns the required similar images set $I_{B}$. 
 \par 
 We model the image similarity as a classification problem with four phases - database formation, feature extraction of images, feature comparison and finally similarity classification (see Fig. 1).
 \vspace{-0.33cm} 
 \begin{figure}[htb]
  \centering
  \includegraphics{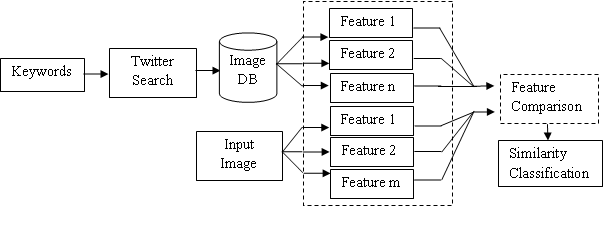}
  \label{fig:f1}
  \vspace{-0.52cm}
   \caption{Proposed methodology to find image similarity. The system takes keywords and an image from the  user,  using the search API of Twitter, creates a database of related images. To compare two images, their features are extracted and fed to a comparison function. Comparison score classifies image similarity.}
\end{figure} 

\vspace{-1.33cm}
\section{Related Work}
\vspace{-0.35cm}

Recent studies show an increase in need of OSM as a plausible resource for first responders \cite{denef2011ict}. Police have realized the effectiveness of OSM in various activities such as investigation, identifying crime, intelligence development, and community policing \cite{denef2011ict,ruddell2013social,solutions2012survey}.
Research shows the effectiveness of OSN during events involving law and order issues like the Boston bombings, Sichuan earthquake (2008), Haiti earthquake (2009), Oklahoma grassfires (2009), and Chile earthquake (2010) \cite{gupta2012misinformation,gupta20131,mendoza2010twitter,starbird2011voluntweeters,vieweg2010microblogging}.
Studies have shown how OSN has been effectively used to aid police to increase community engagement, reduce crime by getting on actionable information from OSN and improve coordination between police and citizen \cite{sachdeva2015online,sachdeva2015social,sachdeva2016social}.  But, these studies focus on the textual aspect of OSN. Research shows that the role of social media text has received attention but the role of social media images remains largely unstudied \cite{kharroub2015social}. However, a report shows that if there is an image attached to a tweet it increases user engagement by 35\% \cite{Rogers_2014}. 
\par
There are also some studies and tools on social media image analysis in general context; like, the analysis of  581 Twitter images  of the 2011 Egyptian revolution revealed that more efficacy-eliciting (crowds, protest activities, national and religious symbols) content is posted by Egyptian users than emotionally arousing (violent) content \cite{kharroub2015social} . Though there have been studies on social media image search, but in most of them query is text based; for instance, a study describes how to use Wikipedia and Flickr content to improve the match of query text with database vocabulary \cite{popescu2011social}. They used a combination of Flickr and Wikipedia query model for query expansion to improve the accuracy of image retrieval systems \cite{popescu2011social}. Images are also used to visually summarise events. In \cite{mcparlane2014picture}, authors propose a technique to find most popular unique images shared on Twitter related to events like sports, law \& politics, art, culture, etc.. In another study, a tool was developed to show trending images of an event by extracting images from Twitter's streaming API using text as query and then detecting near duplicates using locality sensitive hashing \cite{hare2013twitter}. Another hybrid approach for image retrieval combines social relevance by understanding the user's interest using social site Flickr and visual relevance by ranking the images in Google search result according to the interests of the user \cite{cui2014social}. A study shows that image search tools such as TinEye and Google Reverse Image Search are used by journalists to find duplicates, such as other posts of the same image, and near duplicates, such as posts before or after potential Photoshop manipulations, to help find fake posts \cite{wiegand2016veracity}. Though, some of these studies discuss image retrieval aspects but most of them take text as the query to retrieve related images, and systems like TinEye and Google Reverse Image Search gives better results for searches on the entire web than specifically for social media sites like Twitter in specific. Further, these tools do not provide knowledge management to first responders, like users propagating the visual content and sentiments floating with them. Also, most of the image-retrieval tools discussed in these studies use basic hand-crafted features for finding image similarity thus, resulting in comparatively lower accuracy. \par
Our work focuses on emphasizing on the needs of first responders to analyze image spread on social media, for which we deeply analyzed different techniques that can be used for image retrieval with the query as an image and developed a system to aid them to find \& analyze similar images on social media. 
\vspace{-0.45cm}

\section {Data Collection and Annotation} 
\vspace{-0.3cm}
From October, 2015 to February, 2016, we collected data related to 5 events that created law and order issues in society. To create this dataset, we collected data from Twitter using Twitter's search API \footnote[2]{https://dev.twitter.com/rest/public/search}, filtering tweets that contain images, counting to a total of 3,725 images. Fig. 2 shows images which were viral from these events and are taken as input images for evaluating different image similarity models to be studied. The details of the events are discussed below:
\begin{enumerate}
\item Kulkarni Ink: Black ink was sprayed on an Indian technocrat-turned-columnist Sudheendra Kulkarni by the members of a political party, ahead of the launch of former Pakistan's Foreign Minister book, `Neither a Hawk nor a Dove: An In-sider's Account of Pakistan's Foreign Policy'. FIR was lodged against the political party workers and six workers got arrested \cite{kul}. This incidence was slammed on social media arousing political issues and the images of the man with black ink on face went viral. We collected 1,905 tweets with images using the keyword ``\#Kulkarni''. 
\item Baba Ram Rahim: An Indian self-proclaimed Godman Baba Ram Rahim posted pictures posing as Hindu God Vishnu on social media. He was accused by All India Hindu Student Federation for insulting Lord Vishnu and hurting religious sentiments of Hindus by dressing up as Lord Vishnu and lodged a complaint against him \cite{RR}. We collected 408 tweets with images using the keyword ``\#RamRahim''. 
\item Lord Hanuman Cartoon: A cartoon tweeted by an Indian politician drew much flak from other political parties of India and its affiliates for allegedly hurting religious sentiments. The cartoon purportedly showed Hindu Lord Hanuman clad in saffron robes and leader of another political party. The political party lodged a complaint alleging that the cartoon image was posted with the intent to hurt religious sentiments of Hindus \cite{kejri}. We collected 664 tweets with images using keywords like ``\#JNU" and  ``Insults Hanuman''.
\item Charlie Hebdo cartoon: A cartoon in the French satirical magazine Charlie Hebdo sparked outrage by publishing a cartoon attempting to satirize the Syrian refugee crisis. The cartoon imagines Alan Kurdi, the three-year-old Syrian who died in the sea in September 2015, on the way to Europe, has grown up to be a sexual abuser \cite{charlie}. Many called the cartoon was racist and said it was incredibly bad. We collected 568 tweets with images using the keyword ``\#CharlieHebdo''. 
\item Shani Shingnapur protest: As many as 1,500 women, mostly homemakers and college students, planned to storm the Shani Shingnapur temple in India on Republic Day. The protesters wanted to end the age-old humiliating practice of not allowing women to enter the core shrine area \cite{shani}. The image of one of the activist, giving an interview about the protest on Republic Day to media went viral. We collected 180 tweets with images related to the keyword `\#ShaniShingnapur''. 
\end{enumerate}

\begin{figure}[htb]
\centering
\begin{minipage}{.17\linewidth}
\subfigure[]{\label{main:a}\fbox{\includegraphics[height=1.7cm,width=2cm]{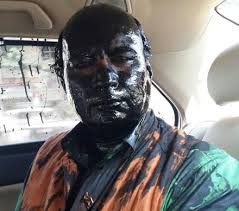}}}
\end{minipage}\hspace{-0.01cm}
\begin{minipage}{.17\linewidth}
\subfigure[]{\label{main:b}\fbox{\includegraphics[height=1.7cm,width=2cm]{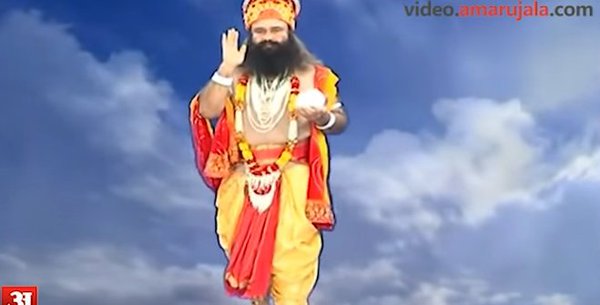}}}
\end{minipage}\hspace{-0.01cm}
\begin{minipage}{.17\linewidth}
\subfigure[]{\label{main:c}\fbox{\includegraphics[height=1.7cm,width=2cm]{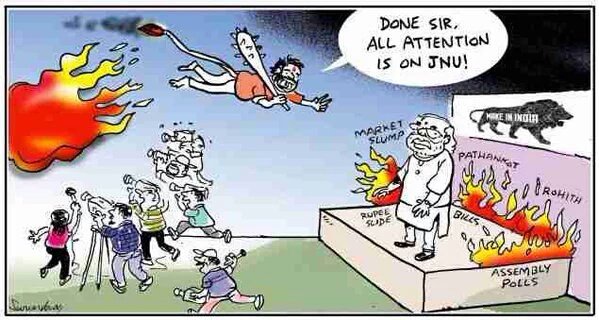}}}
\end{minipage}\hspace{-0.01cm}
\begin{minipage}{.17\linewidth}
\subfigure[]{\label{main:d}\fbox{\includegraphics[height=1.7cm,width=2cm]{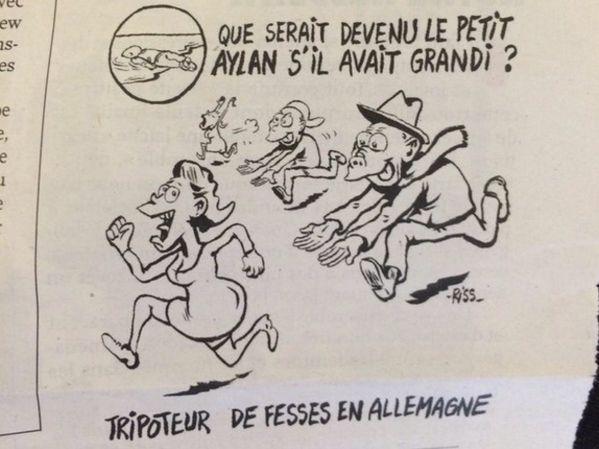}}}
\end{minipage}\hspace{-0.01cm}
\begin{minipage}{.18\linewidth}
\subfigure[]{\label{main:e}\fbox{\includegraphics[height=1.7cm,width=2cm]{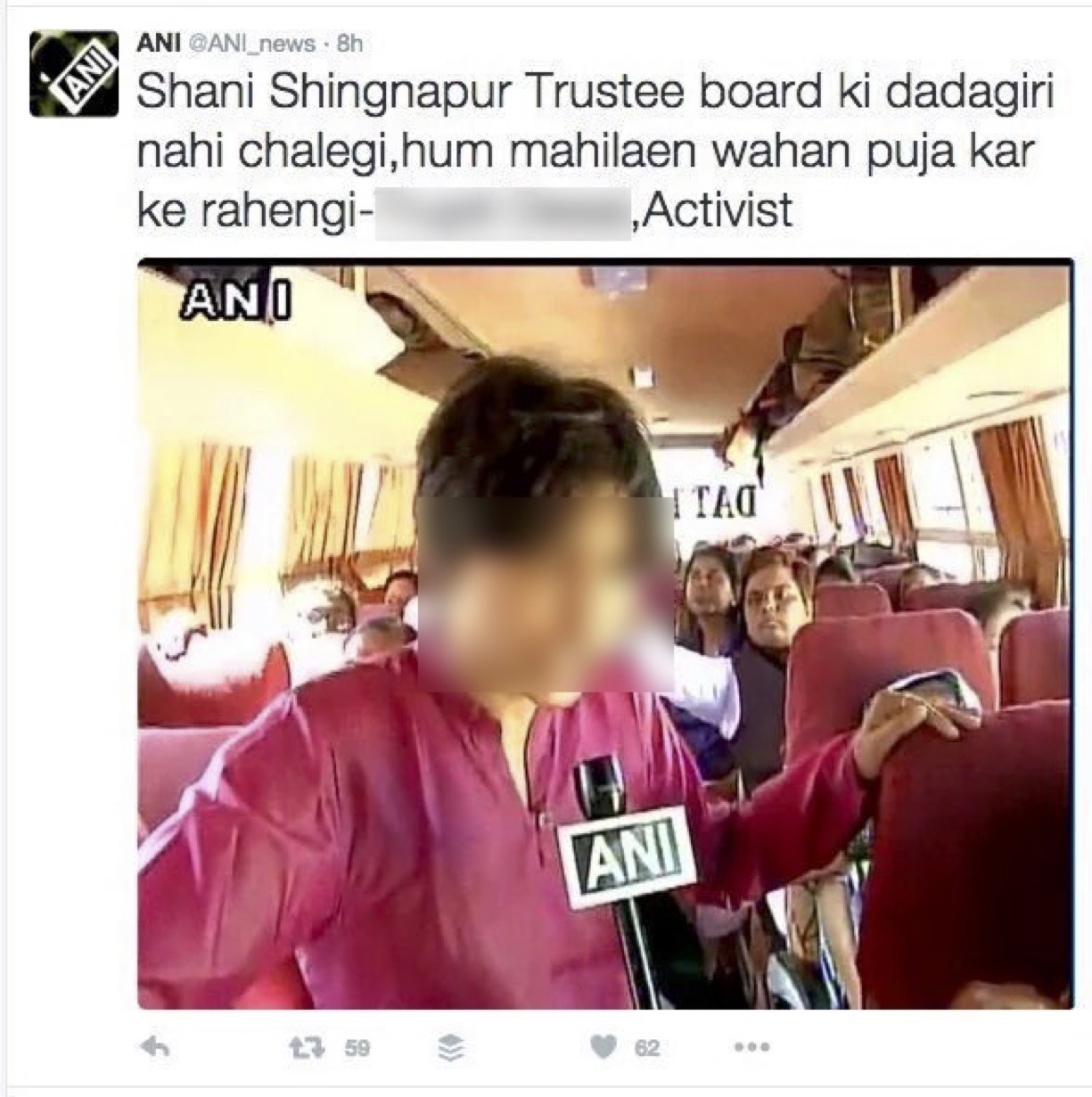}}}
\end{minipage}
\vspace{-0.2cm}
\caption{Input images set. These images represents image which could be of interest to first responders and were viral from the event (a) Kulkarni Ink (b) Baba Ram Rahim (c) Lord Hanuman cartoon (d) Charlie Hebdo cartoon (e) Shani Shingnapur, respectively.}\label{fig:main}
\end{figure}
\textbf{Data Annotation}
\newline
To create the ground truth for testing the models, we obtained labels for all the 3,725 images. The annotators were given an image from each event (see Fig. 2) and from the respective event's image database they had to mark images that were similar and dissimilar to the given image. The broad definition of image similarity given to them was: images that are cropped, scaled, blurred, wrapped with text, stitched with other images, having color changes, brightness changes or contrast enhancements are considered similar. The images which had agreement from at least two annotators were marked in the final results. Table 1 shows the dataset after annotation was completed.

\begin{table}[b]
\centering

\begin{tabular}{ |c|c|c|c| }
 \hline

Keyword &	Total Images   &	Similar Images  &	Dissimilar Images \\
 \hline
\#Kulkarni 	& 1,905 &	354 &	1,551 \\
\#RamRahim &	408 & 	97 & 	311 \\
\#Insults Hanuman &	664 &	277 &	387 \\
\#CharlieHebdo	 &568 &	118	&450\\
\#ShaniShingnapur &	180 &	70 &	110 \\
 \hline
\end{tabular}
\label{Table:t1}
\vspace{0.1cm}
\caption{Data after annotation shows the number of similar and dissimilar images for each event.}
\end{table}
\vspace{-0.55cm}
\section {Similarity Modelling}
 \vspace{-0.33cm}
The retrieval performance of CBIR (Content Based Image Retrieval) system crucially depends on the features that represent images and their similarity measurement. In this section, we'll discuss some of the techniques we studied to compare the similarity of two images. \vspace{-0.4cm}
\subsection {Hand-Crafted Features}
 \vspace{-0.2cm}
\begin{enumerate}
\item Color-based feature similarity \par
Color histograms are frequently used to compare images \cite{pass1997comparing}. It is often done by comparing color histograms of images, which eliminates information on the spatial distribution of colors. The image descriptor is a 3D RGB color histogram with 8 bins per channel and we compare the descriptor using a similarity metric. After extracting image descriptors, we calculated Bhattacharyya distance for comparing the two image's histogram descriptors. The lesser is the distance between two color histograms, more similar the two images are. 
\item Keypoint-Descriptor-based similarity \par
Certain parts of an image have more information than others, particularly at edges and corners. Keypoints can be generated using this information. After finding the keypoints of an image, next step is to find their descriptors. Descriptors are fixed length vectors that describe some characteristics about the keypoints. Next, we compare each keypoint descriptors of one image to each keypoint descriptors of the other image. Since, the descriptors are vectors of numbers, we can compare them using different distance metrics. We studied techniques like DAISY \cite{tola2008fast}, ORB \cite{rublee2011orb} and Improved ORB \cite{yu2015improved} for computing keypoint descriptors and then comparing them.  
\let\labelitemi\labelitemii

\begin{itemize}
\item 
The \emph{DAISY} dense image descriptor is based on gradient orientation histograms similar to the SIFT (Scale Invariant Feature Transform) descriptor. It is formulated in a way that allows for fast dense extraction which is useful for bag-of-features image representations \cite{tola2010daisy}. After extracting DAISY descriptors we calculate the distance between the descriptor vectors of two images using Brute Force matcher with KNN (K-Nearest Neighbors) as the distance metric. We use this distance as the score to measure the similarity of two images. 

\item 
\emph{ORB} uses improved FAST for feature detection, and these features are described using an improved Rotated BRIEF feature descriptor \cite{yu2015improved}. Since the speed of FAST and BRIEF are very fast this can be the choice for real-time systems. ORB is rotational invariant, noise invariant and uses image pyramids for scale invariance \cite{rublee2011orb}. It returns binary strings to describe feature points, which is used for feature point matching using Hamming distance. We use this hamming distance as the score to evaluate the similarity of two images. 
\item 
In order to improve the matches given by ORB we add one extra step. After comparing the keypoint descriptors given by ORB, we find homography of two images using RANSAC \cite{yu2015improved}. We extract top 30 matches having the least distance and pass these matches to RANSAC to weed out wrong matching points. This technique is called as \emph{Improved ORB}. RANSAC returns a binary array equal to size of input matches array, where 0 represents a false match and 1 represents a true match. We define true ratio as the ratio of true matches returned by RANSAC to total matches given to it. We use this true ratio to evaluate the similarity of two images. In the given equation, $t_{r}$ denotes true ratio, $n$ is the size of matches set passed to RANSAC, $A$ is the array returned by RANSAC and $A_{i}$ represents the value of array returned by RANSAC at $i^{th}$ index. 

$$t_{r}= \frac{\sum_{i=0}^{n-1} A_{i}}{n} $$

\end{itemize}
\end{enumerate}
\vspace{-0.75cm}
\subsection{Trained Features}
\vspace{-0.3cm}
In this section we discuss one important technique ``deep learning'' which is organised in a deep architecture and processes information through multiple stages. To that end, inspired by recent advances in neural architectures and deep learning, we choose to address CBIR using deep convolutional neural network.
\par
For implementation of \emph {Convolutional Neural Network (CNN)}, we took inspiration from the architecture discussed in \cite{radford2015unsupervised}. To create the dataset for training, we used an unsupervised approach. We extracted around 25K images from popular news websites and then on each of 25K image, we applied various image processing operations like cropping, distorting, blurring, scaling, adding text, stitching image, adding noise, and changing color, to generate the corresponding similar image. Thus, we created a dataset of 50K images. To train the model we had 25K pairs of similar and dissimilar images. Similar approach of transforming images to create dataset was used by Fischer et al., 2014 \cite{fischer2014descriptor}. As a pre-processing step we first align the images.
We extract SIFT descriptors to map the matching features of both the images and align the images by applying a perspective transformation using the homography. We then club the two aligned images to form a 6 channel image which is passed to the CNN model. \par
The model contains 15 layers out of which 6 are trainable. First 5 of trainable layers are convolutional and one is fully connected. The output of the last fully connected layer is a 1D vector and is fed to a sigmoid activation which produces the final output between 0 and 1. Our network minimizes binary cross entropy loss objective and uses adam optimizer as the optimizer. The convolutional layers contain 64 filters $122\times122$ pixels, 32 filters $61\times61$ pixels, 16 filters $31\times31$ pixels, 6 filters $16\times16$ pixels, 1 filter $16\times16$ pixels.  All the convolution layers are applied with a stride of 4 pixels.  The fully connected layer contains 1 neuron. All the convolution layers use LeakyRelu and batch normalization except the first layer which doesn't use batch normalization and the last convolution layer directly feeds output to the fully connected layer. All the layers use a subsampling of $2\times2$ except the last layer, which uses a subsampling of $1\times1$. The initial convolution layers extract the low level features like edges and gradients. The other convolution layers learn to compare the extracted features of the two images. The similar areas of the images get propagated till top and fully connected layers calculate a similarity score using the high level features. 

\vspace{-0.45cm}
\section{Hand-crafted vs. Trainable Features}
\vspace{-0.33cm}
In this section, we evaluate different techniques discussed in the above section on the annotated test data described in section 4. 
To evaluate the performance of the above-discussed image similarity models we calculate accuracy of their classification results, which is the ratio of true classification to the total population. 
\vspace{-0.9cm}
\subsection{Histogram}
\vspace{-0.3cm}
For each of the five events, we plotted the accuracy for Bhattacharyya distance between color histogram of two images, ranging from 0.1 to 0.9. Fig. 3 (a) shows accuracy for different events at different histogram distance. The graph shows a lot of variation in the accuracy corresponding to the histogram distance values. We calculated average variance in accuracy at 3 distance points on the x-axis (0.2, 0.4, 0.5) as 104.24, which is high making it tough to choose a distance value as the threshold. For example, If we choose threshold distance as 0.4, the accuracy for ShaniShignapur and Lord Hanuman cartoon is close to 60\% whereas for Charlie Hebdo it is 90\%. Another drawback of this method is, it lacks spatial information, so images with very different appearances can also have similar histogram \cite{pass1997comparing}. 
 \vspace{-0.4cm}
\subsection{DAISY}
\vspace{-0.3cm}
The mean distance between DAISY descriptors of  both similar and dissimilar images was in the range of 0.0 to 0.10. To choose the threshold distance values between this range, we plotted the accuracy graph for each event at different threshold value. Fig. 3(b) shows the accuracy vs. threshold plot, where the threshold is the mean distance between DAISY descriptors of two images. Since the range of distance returned by DAISY descriptors is very small there was a very high overlap between distance values of similar and dissimilar images leading to high error. Due to this high error, the maximum accuracy achieved overall is less than 88\%, at a distance of 0.06 but, at this point accuracy for remaining events 4 events is less than 50\%. We also calculated the average variance in accuracy at 3 distance points on the x-axis (0.02, 0.05, 0.06) to be 196.58, which is even higher than the Histogram technique. Thus, making it hard to choose an optimal threshold value.   
\vspace{-0.5cm}
\begin{figure}[htb]
\subfigure[]{%
  {\includegraphics[height=4cm,width=6cm]{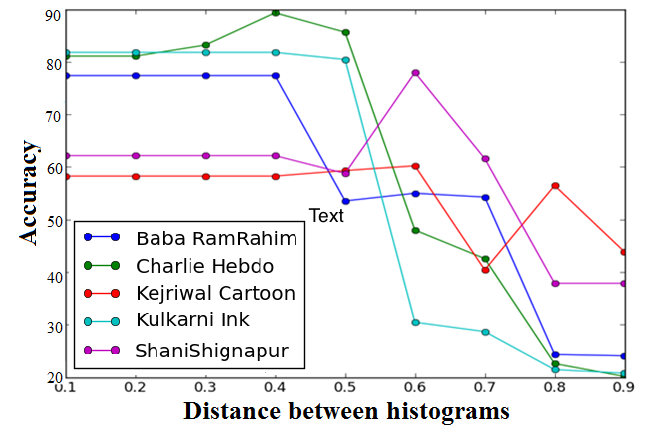}\label{fig:f5}}}
\subfigure[]{%
{\includegraphics[height =3.9cm,width=5.9cm]{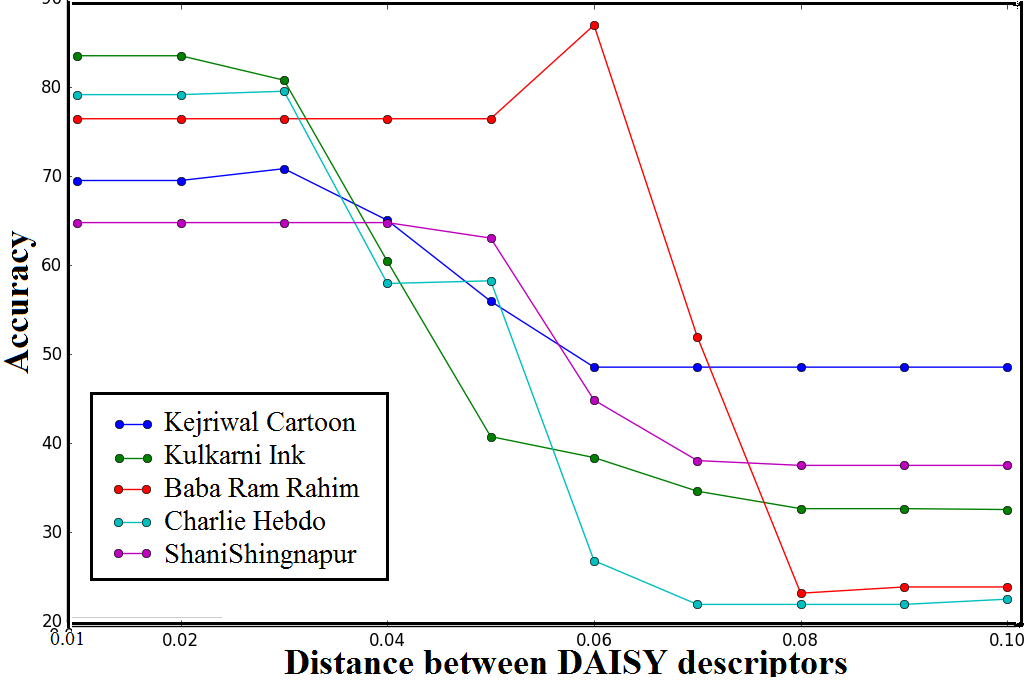}\label{fig:f8}}}
\vspace{-0.4cm}
 \caption{ (a) Accuracy for different events at different distance between histograms of two images. (b) Accuracy for different events for different distance between DAISY descriptors of two images. Both graphs shows high variance in accuracy.}
\end{figure}
\vspace{-1.2cm}
\subsection{ORB}
\vspace{-0.3cm}
For all the 5 events, we plotted the accuracy for different values of mean distance between ORB descriptors. We use this plot to choose an optimal threshold distance value. Fig. 4(a) represents the accuracy plot for different events at different distance between ORB descriptors. The variation in accuracy is less as compared to the Histogram and DAISY technique. Average variance in accuracy at 3 distance points on the x-axis (29, 32, 35) is 17.6. Also, the accuracy for all the events at threshold distance 29 is above 88\%. This shows that ORB is certainly a better choice than Histogram and DAISY for our data.
\vspace{-0.5cm}
\subsection{Improved ORB}
\vspace{-0.3cm}
In Improved ORB, after getting the match set of the descriptors from ORB, we pass the top thirty matches having least distance to RANSAC, which filters matches which are true. The threshold value here is the ratio of true matches returned by RANSAC to the total matches passed \big(thirty in our case\big), defined as true match ratio. Fig. 4(b) shows the accuracy of Improved ORB for different true match ratio taken as the threshold. Average variance in accuracy at 3 true match ratio values on the x-axis (0.33, 0.35, 0.37) is 6.2, which is least among all the techniques discussed till now.  Also, after true match ratio of 0.3, the accuracy for each event is almost constant and is above 90\% for all events. If compared with above methods Improved ORB is giving best results. Hence, Improved ORB can be termed the state-of-art technique in hand-crafted features discussed.

\begin{figure}[htb]
 \subfigure[]{%
{\includegraphics[height= 4cm,width=0.5\textwidth]{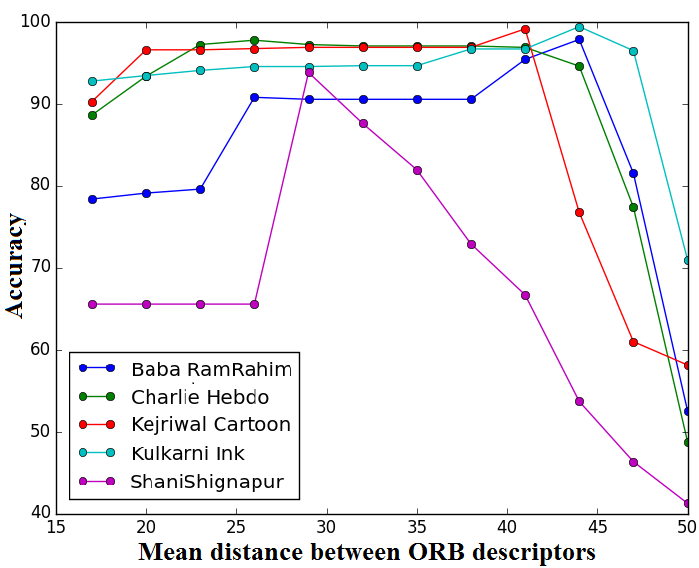}\label{fig:f9}}}
\subfigure[]{%
{\includegraphics[height= 4cm,width=0.5\textwidth]{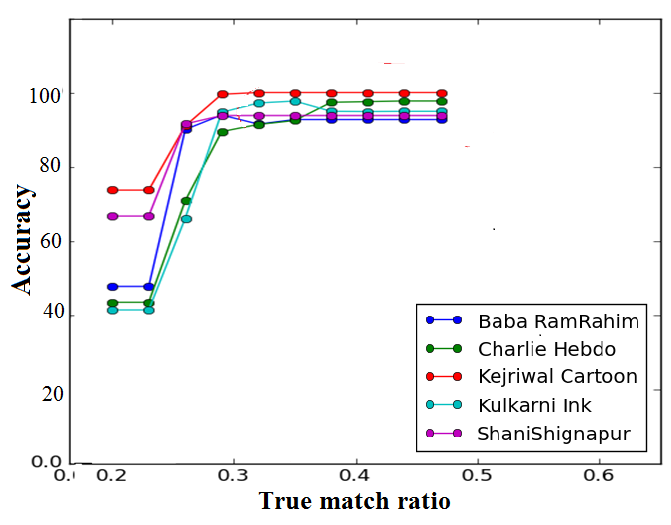}\label{fig:f10}}}
\vspace{-0.4cm}
\caption[ORB and Improved ORB]{(a) Accuracy for different events at different distance values between ORB descriptors, at distance value 29 on x-axis all the events have accuracy more than 88\% and (b) Accuracy for different events at different true match ratio returned by Improved ORB, at true ratio of 0.35 all events have accuracy above 90\%.}
\end{figure}
\vspace{-1.1cm}
\subsection{Trainable Features}
\vspace{-0.3cm}
We trained a deep CNN model for 25K pairs of similar and dissimilar images, and to test the model we gave input image and images from annotated set to the model for different epoch values. To choose optimal epoch for the model, we plot the accuracy for different events, for models trained on different epoch values. Fig. 5 shows the accuracy for each event at different epoch value. The model trained for epoch value 35 is showing an accuracy above 97\% for all events. Also, average variance at three epoch values on the x-axis (30, 35, 40) is 0.71, which is even lower than what we achieved in Improved ORB. Since the overall accuracy for CNN model is higher and the variance is lesser than Improved ORB, we can conclude that our proposed CNN model outperforms state-of-art Improved ORB. In the next section, we'll see how these two techniques perform when the input image is modified. 
\vspace{-0.5cm}
\begin{figure}[htb]  
 \centering
  {\includegraphics[height=3.8cm,width=0.5\textwidth]{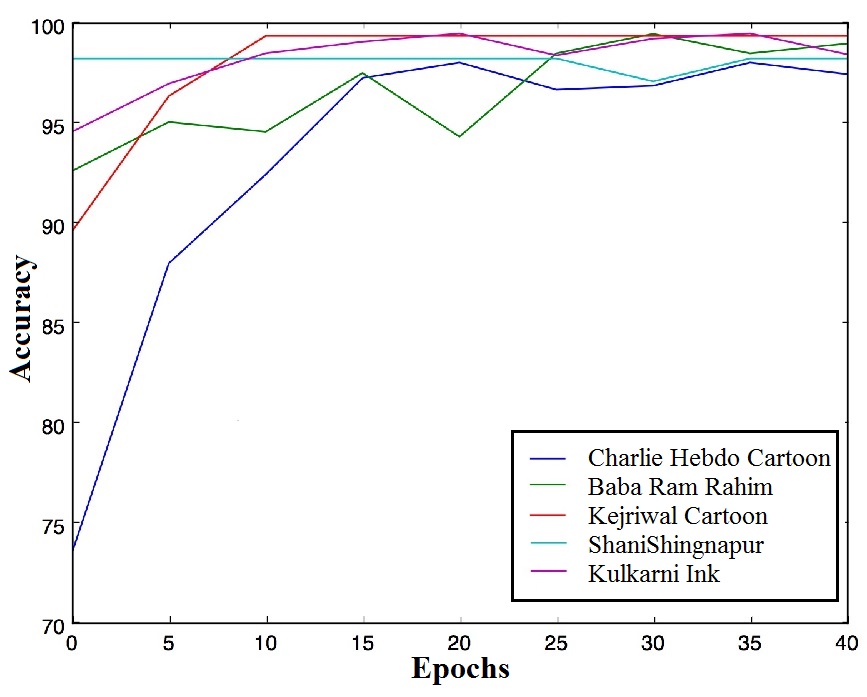}\label{fig:f10}}
  \vspace{-0.32cm}
  \caption[CNN]{CNN models trained for different epochs and their accuracy for five test events. For model trained for 35 epochs, accuracy for all events is above 97\%.}
\end{figure}
\vspace{-1.3cm}
\section{Competing on Modified Image}
\vspace{-0.35cm}
In this section, we analyze the performance of Improved ORB (state-of-art technique in hand-crafted features) and CNN when the input image is modified. For all the 5 events, we take different cases of modified images usually seen on social media sites like Twitter and compare the accuracy of both the image retrieval techniques. Fig. 6 shows modified images taken from all the events and Table 2 shows the accuracy of the image retrieval models for these images. For each event, we picked three kinds of modified images: one with single modification like cropped or scaled or color changes, second with dual modification together like cropping \& scaling or adding text \& scaling, etc., and third with multiple modifications together like scaling, cropping, adding text and stitching image, represented as All in Table 2.  \par 
Among different modifications seen, the most common modification technique was scaling the image. Hence, we calculated the scaling factors of the modified images and found that as the scaling factor increases, the accuracy dip for Improved ORB also increases. For example, in Table 2 image `a' is modified by a scaling factor of $7.42\times5.23$ and the accuracy of Improved ORB is 84.1\% while, for the same image CNN is showing an accuracy of 99.4\%. Likewise, for image  `c', `m', `n', `o' also the scaling factor is high leading to low accuracy in Improved ORB but, the accuracy of CNN in all these cases outperforms Improved ORB. 
Another common modification affecting accuracy is cropping of an image, for instance; image `f' has only the cropped face from the complete body of Baba Ram Rahim (original picture see in Fig. 2), and again the accuracy dip in Improved ORB is much higher than CNN. Likewise, is the case for image `d', cropping reduced the accuracy of Improved ORB while CNN still outperforms it.  Hence, we can conclude that the proposed CNN model is more robust to modifications like high scaling and high cropping than state-of-art Improved ORB.  

\begin{figure}
\centering
\hspace{-0.2cm}
\minipage{0.27\textwidth}
\subfigure[]{\label{main:a}\fbox{\includegraphics[height=1.5cm,width=3cm]{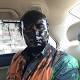}} }
\endminipage
\hspace{-0.1cm}
\minipage{0.27\textwidth}
 \subfigure[]{\label{main:2}\fbox{\includegraphics[height=1.5cm,width=3cm]{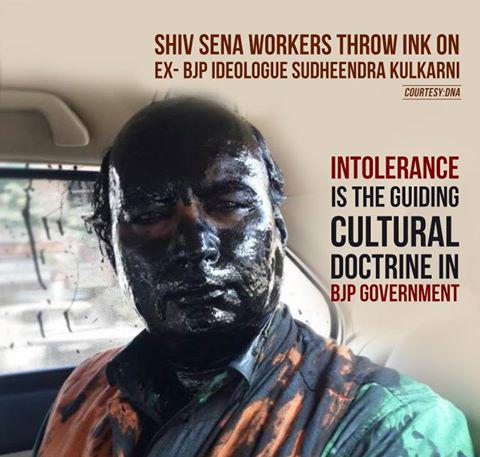}}}
 \endminipage
 \hspace{-0.25cm}
 \minipage{0.27\textwidth}%
\subfigure[]{\label{main:3} \fbox{ \includegraphics[height=1.5cm,width=3cm]{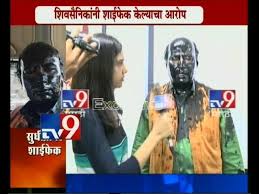}}}
 \endminipage  \\
\minipage{0.27\textwidth}
\subfigure[]{\label{main:4}\fbox{  \includegraphics[height=1.5cm,width=3cm]{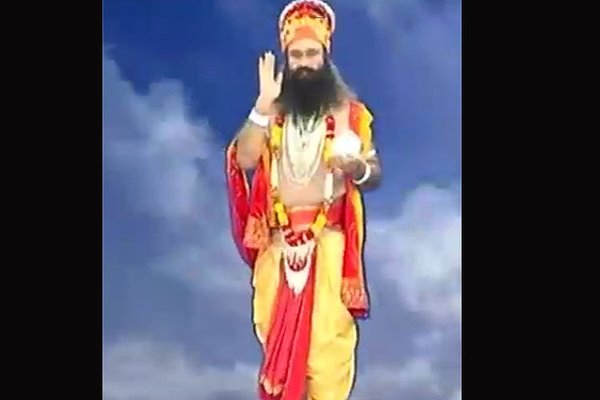}}}
 \endminipage
 \minipage{0.27\textwidth}
 \subfigure[]{\label{main:5}\fbox{\includegraphics[height=1.5cm,width=3cm]{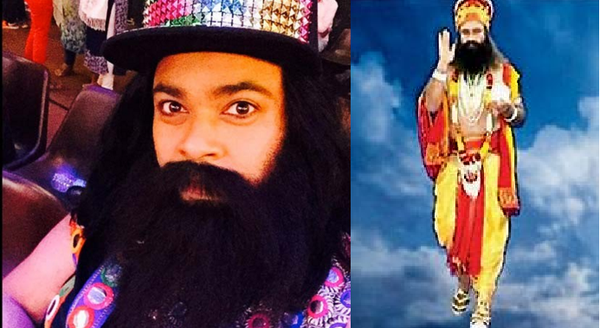}}}
\endminipage
\minipage{0.27\textwidth}%
 \subfigure[]{\label{main:6}\fbox{\includegraphics[height=1.5cm,width=3cm]{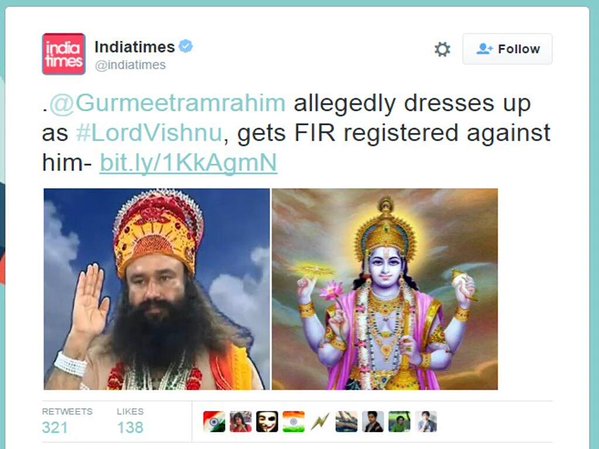}}}
\endminipage \\
\hspace{-0.2cm}
\minipage{0.27\textwidth}
 \subfigure[]{\label{main:7}\fbox{\includegraphics[height=1.5cm,width=3cm]{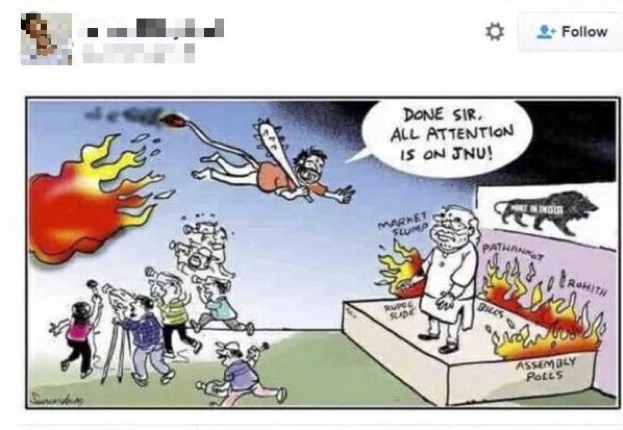}}}
\endminipage
\hspace{-0.22cm}
\minipage{0.27\textwidth}
\subfigure[]{\label{main:8} \fbox{\includegraphics[height=1.5cm,width=3cm]{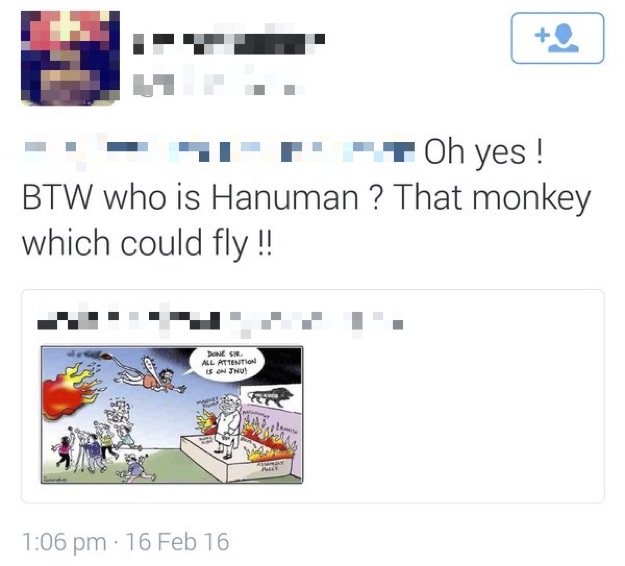}}}
\endminipage
\minipage{0.27\textwidth}%
 \subfigure[]{\label{main:9}\fbox{\includegraphics[height=1.5cm,width=3cm]{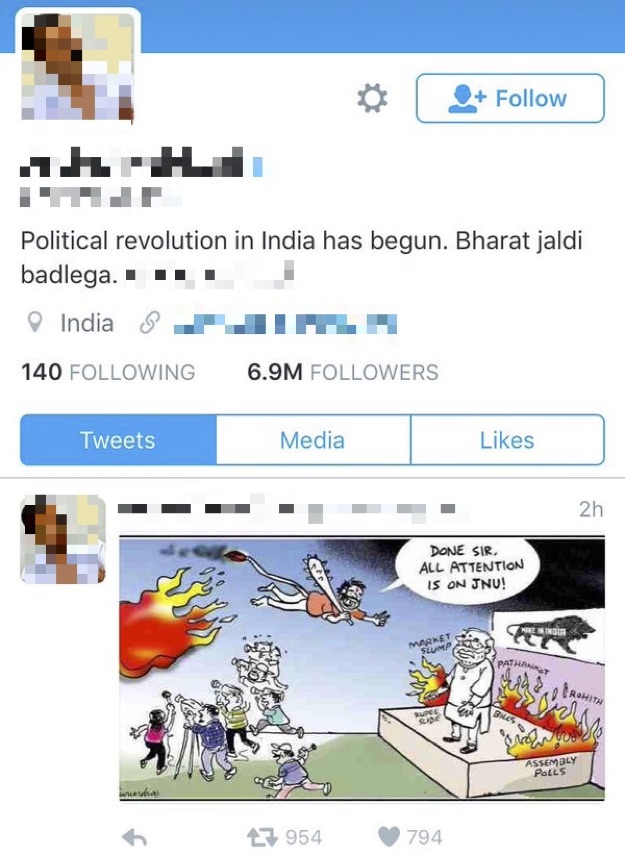}}}
\endminipage \\
\minipage{0.27\textwidth}
 \subfigure[]{\label{main:10}\fbox{\includegraphics[height=1.5cm,width=3cm]{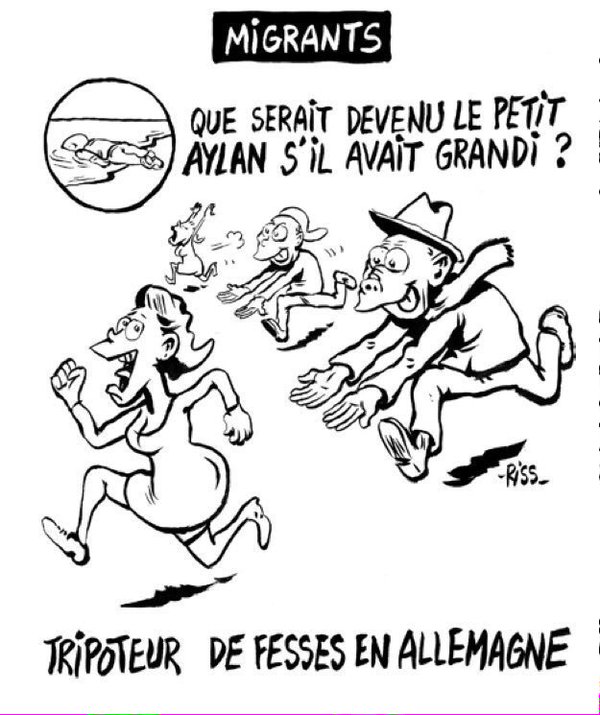}}}
\endminipage
\minipage{0.27\textwidth}
\subfigure[]{\label{main:11}\fbox{ \includegraphics[height=1.5cm,width=3cm]{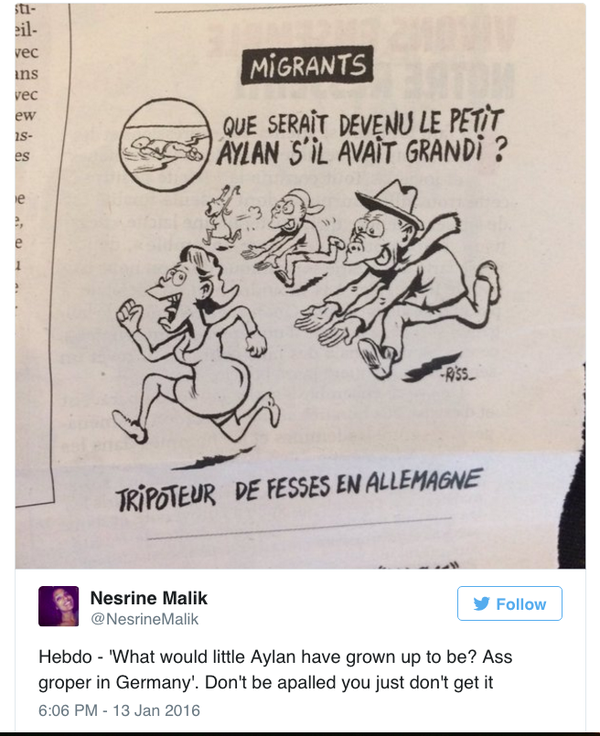}}}
\endminipage
\minipage{0.27\textwidth}%
 \subfigure[]{\label{main:12}\fbox{\includegraphics[height=1.5cm,width=3cm]{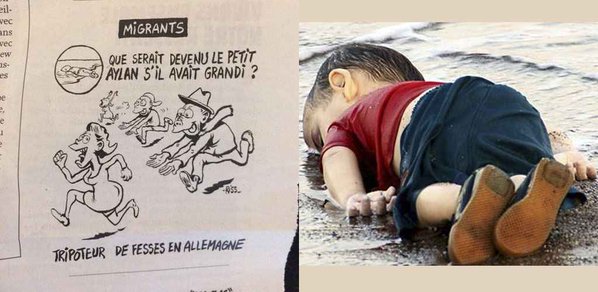}}}
\endminipage \\
\minipage{0.27\textwidth}
 \subfigure[]{\label{main:13}\fbox{\includegraphics[height=1.5cm,width=3cm]{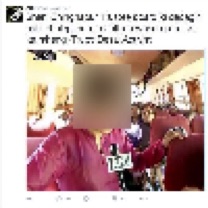}}}
\endminipage
\minipage{0.27\textwidth}
 \subfigure[]{\label{main:14}\fbox{\includegraphics[height=1.5cm,width=3cm]{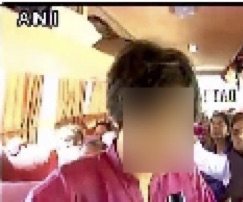}}}
\endminipage
\minipage{0.27\textwidth}
 \subfigure[]{\label{main:14}\fbox{\includegraphics[height=1.5cm,width=3cm]{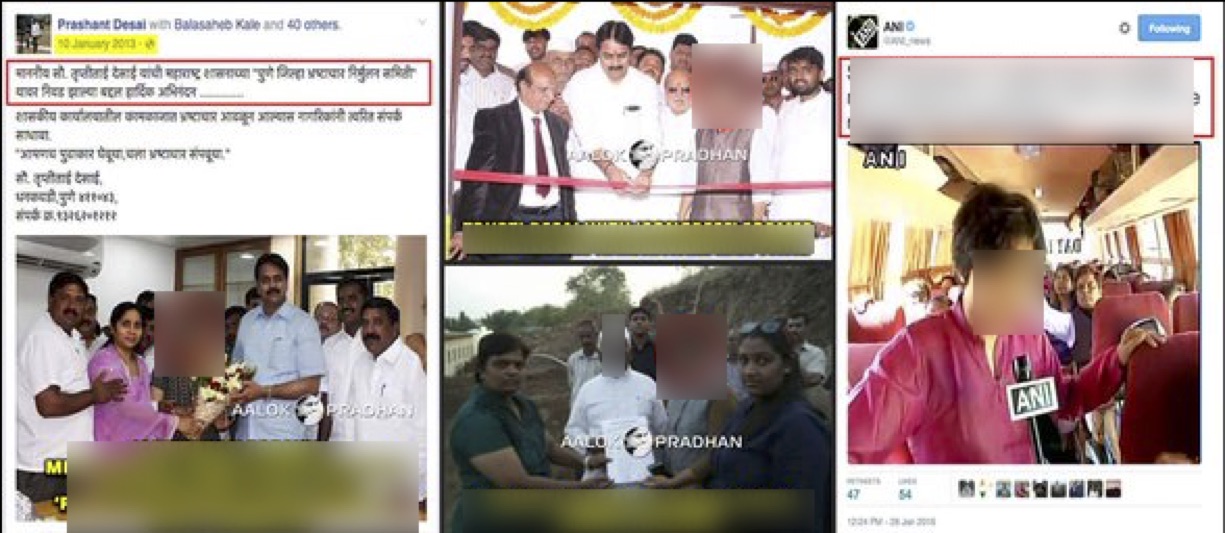}}}
\endminipage

\centering
\vspace{-0.15cm}
\caption{Event-wise modified images. By Row: (1) Kulkarni Ink Event (2) Baba Ram Rahim (3) Lord Hanuman Cartoon (4) Charlie Hebdo Cartoon (5) Shani Shingnapur.} 
\vspace{0.4cm}

\begin{minipage}[b]{1.0\linewidth}
\centering
\begin{tabular}[b]{ |c|c|c|c|c|c|}
\hline
Image-Id & Improved ORB & CNN & Modification & Scaling Factor & Event\\
\hline

(a) & 84.1 & 99.4 & Scaled & $7.42\times5.23$ & Kulkarni Ink \\ 
(b) & 95.6 & 99.5 & Text added \& scaled  &  $1.23\times1.08$ & Kulkarni Ink \\
(c) & 80.7 &  87.3 & All & $9.61\times 3.87$ & Kulkarni Ink \\

(d) & 91.2 &  99.5 & Cropped & -- & Baba Ram Rahim \\
(e) & 88.2 &  92.6 & Stitched \& scaled & $2.31\times0.92$ & Baba Ram Rahim \\
(f) & 77.4 &  93.8 & All & $2.65\times1.56$ & Baba Ram Rahim \\

(g)& 96.9 &  99.3 & Scaled  & $1.04\times1.09$ &Lord Hanuman Cartoon \\
(h) & 90.3 &  99.5 & Text added \& scaled & $2.32\times2.47$ & Lord Hanuman Cartoon \\
(i) & 92.2 &  92.7 & All  & $1.27\times1.25$ & Lord Hanuman Cartoon \\

(j) & 96.8 &  97.2 & Background color & -- &Charlie Hebdo Cartoon \\
(k) & 97.2 &  97.6 & Text added \& scaled & $1.04\times1.28$ & Charlie Hebdo Cartoon \\
(l) & 94.5 &  95.3 &  All  & $2.02\times1.54$ & Charlie Hebdo Cartoon \\

(m) & 90.4 &  95.3 & Scaled & $4.98\times6.09$ & Shani Shingnapur \\
(n) & 62.1 &  98.2 & Cropped \& scaled &  $5.88\times5.91$ &Shani Shingnapur \\
(o) & 93.7 &  98.3 & All & $3.24\times 2.35$ & Shani Shingnapur \\

\hline
\end{tabular}
\end{minipage}
\captionof{table}{ Table showing the accuracy comparison of different modified images for test events. In all the test cases, CNN has out-performed state-of-art Improved ORB.}
\end{figure}
\vspace{-0.5cm}
\section{Benefits: Improved Understanding of Law \& Order Scenario}
\vspace{-0.5cm}
The system is built with the aim to aid first responders to find the spread of images that can create law and order situations by retrieving similar images, find \& analyze the users propagating the content, sentiments floating, etc. Thus, helping in reducing their human efforts in identifying the visual content of the interest. In the previous sections, we saw comparison between different hand-crafted methods and CNN to compare image similarity. The experimental results show that CNN outperforms Improved ORB and hence, CNN is a more suitable technique to find image spread for the system. We now explain various other measures offered by the system that help in achieving goals for better law enforcement using similar image retrieval process discussed above. \par 
\textbf{Search Space Reduction}: The system takes a set of keywords from the user to create a database of images and an image using which it retrieves a set similar images from the database created. Thus, reducing the search space for first responders hence, saving their time and efforts to analyse the overall dump. We were able to reduce the search space on an average by 67\% and by 65\%, using CNN and Improved ORB as image retrieval technique respectively. The maximum reduction in search space was seen for Baba Ram Rahim event followed by Charlie Hebdo cartoon. Table 3 shows the details of search space reduction for all the events using the two techniques.  \par \textbf{Users Analysis}: Users play a vital role in spreading the content on social media. Thus, it is important for security analysts to identify people who are spreading the images. The system tries to deliver this need by producing a list of users who spread the visual content and their details like, their Twitter usernames, profile pictures, description, location, and links to their Twitter profiles. In our dataset, we found maximum users for Lord Hanuman cartoon event propagating the content, for images retrieved using both CNN and Improved ORB. Table 3 shows the number of users listed when we used Improved ORB and CNN as image retrieval techniques. \par \textbf{Sentiment Analysis \&  Tweets vs. Retweets Analysis}: Sentiments of the tweets play an important role in spreading the content and eliciting the reaction of people. Thus, making it an important analysis to find the percentage of content having positive, negative and neutral sentiments. The system currently analyzes the sentiment of the textual content having images using Sentiment140 API \footnote[3]{http://help.sentiment140.com/home}, widely used in the literature to study social media data. In our dataset, we found 7\% of tweets have negative sentiments and 5\% have positive sentiments and remaining have neutral sentiments which might be due to broken language, non-english, hinglish language  or absence of text. Tweets vs. Retweets analysis shows the percent of retweets and original tweets. The analysis shows, most of the images in the result were spread by retweeting, average retweeted posts were 89.4\%. 
\vspace{-0.6cm}
\begin{table}[]
\centering
\begin{minipage}[htb]{4.3in}
\centering
\begin{tabular}{|c|c|c|c|c|}
\hline
\multirow{2}{*}{Event} & \multicolumn{2}{c|}{\begin{tabular}[c]{@{}c@{}}Search Space Reduction (\%)\end{tabular}} & \multicolumn{2}{c|}{\begin{tabular}[c]{@{}c@{}}Number of Users\end{tabular}} \\ \cline{2-5} 
                       & Improved ORB                                     & CNN                                      & Improved ORB                                & CNN                               \\ \hline
Kulkarni ink           & 51.1                                             & 54.3                                     & 770                                         & 1,144                                 \\ \hline
Baba Ram Rahim              & 83.1                                             & 90.7                                     & 75                                          & 96                                \\ \hline
Lord Hanuman cartoon              & 60.1                                             & 61.6                                     & 2,210                                       & 2,657                                \\ \hline
Charlie Hebdo cartoon               & 74.1                                             & 75.2                                     & 2,132                                       & 2,037                                \\ \hline
Shani Shingnapur                   & 57.9                                             & 57.9                                     & 186                                         & 186                               \\ \hline
\end{tabular}
\vspace{0.06cm}
\caption{Event-wise reduction in search space and analysis on number of users after retrieving similar images using Improved ORB and CNN.}
\label{my-label}
\end{minipage}
\end{table}
\vspace{-1.9cm}
\section {Implementation and Response Time}
\vspace{-0.55cm}
\subsection{Implementation}
\vspace{-0.55cm}
The proposed system takes an image and keywords as input. These keywords are then given to Twitter's Search API (Application Program Interface), part of Twitter's REST API. The API returns a set of tweets related to the keywords it takes as the query, and the system then filters and saves the tweets containing images in a database. As the images get stored in the database, the image comparison model computes the similarity score between the input image and the images in the database. We then set a threshold $t$, if the similarity score is above $t$, images are marked similar else dissimilar. After getting similar images, the system does analysis on the retrieved set of images and finds the users propagating them, the sentiment analysis and retweets analysis. Fig. 7 shows input and different output and analysis screens of the proposed system. 
\begin{figure}[htb]
            {\fbox{{\includegraphics[ height= 4.8cm,width=\textwidth]{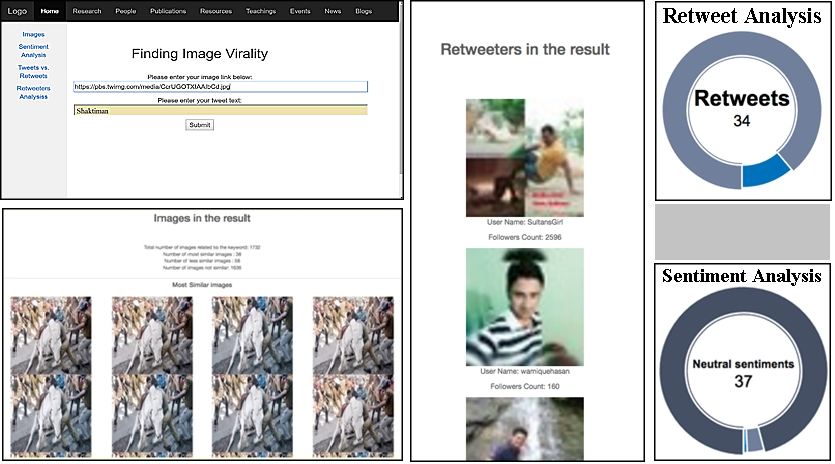}}}}
   \vspace{-0.4cm}
    \caption [Screen Shot of the system.]{Moving in anti-clockwise from top-left image the images represent (a)Input to the system is a image url and a keyword.  (b) Output screen: The system retrieves similar images (c) Analysis screen: Users propagating (d) Analysis screen: Sentiment analysis on text of the resultant images (e) Analysis Screen: Retweets analysis on text of the resultant images.}
    \label{figure:system-input-output}
\end{figure} 
\vspace{-0.45cm}
\subsection{Response Time}
\vspace{-0.2cm}
Since our aim is to build a real-time image search system, apart from the accuracy of the model we also need to select the technique that is time efficient. We compare the approximate time consumed by Improved ORB and CNN to compare one pair of image and we found that Improved ORB took 0.2 seconds and CNN model took 0.55 seconds.
\vspace{-0.45cm}
\section{Conclusion}
\vspace{-0.35cm}
In our study, we proposed a OSINT real-time image search system robust to find modified images, that aids first responders to analyze and find the current spread of images creating law \& order situations in society, analyze users propagating such images and sentiments floating with them. 
The system performs a hybrid search by taking a keyword to create an image database and an image as input and returns similar images. The system also aims to reduce search space for first responders on average by 67\%.
To find similar images we conducted an in-depth experimental analysis of different CBIR techniques. We did the experimental study on the data collected for the 5 events which created law and order situations in society and whose images were viral during the duration of the study. We compared different hand-crafted techniques and found that Improved ORB (ORB + RANSAC) is the state-of-art technique for the hand-crafted approach. We also proposed a CNN model which outperforms the accuracy of state-of-art hand-crafted technique. Future work will focus on reducing the response time of the proposed system, improving results of textual sentiment analysis that takes broken or non-english language into consideration, including more image analytics features like, sentiments of images in result to aid maintaining safety in society during security critical scenarios. 
\newline
\newline
\bibliographystyle{plain}
\bibliography{Bibliography}

\begin{thebibliography}{10}

\bibitem{crump2011police}
Jeremy Crump.
\newblock What are the police doing on twitter? social media, the police and
  the public.
\newblock 3(4):1--27, 2011.

\bibitem{cui2014social}
Peng Cui, Shao-Wei Liu, Wen-Wu Zhu, Huan-Bo Luan, Tat-Seng Chua, and Shi-Qiang
  Yang.
\newblock Social-sensed image search.
\newblock {\em ACM Transactions on Information Systems (TOIS)}, 32(2):8, 2014.

\bibitem{denef2011ict}
S~Denef, N~Kaptein, and PS~Bayerl.
\newblock Ict trends in european policing. composite project, 2011.

\bibitem{shani}
The~Indian Express.
\newblock Over 1500 women plan to storm shani shingnapur temple today, January
  2016.

\bibitem{fischer2014descriptor}
Philipp Fischer, Alexey Dosovitskiy, and Thomas Brox.
\newblock Descriptor matching with convolutional neural networks: a comparison
  to sift.
\newblock {\em arXiv preprint arXiv:1405.5769}, 2014.

\bibitem{gerber2014predicting}
Matthew~S Gerber.
\newblock Predicting crime using twitter and kernel density estimation.
\newblock {\em Decision Support Systems}, 61:115--125, 2014.

\bibitem{charlie}
The Guardian.
\newblock Charlie hebdo cartoon depicting drowned child alan kurdi sparks
  racism debate, January 2016.

\bibitem{gupta2012misinformation}
A~Gupta and P~Kumaraguru.
\newblock Misinformation on twitter during crisis events. encyclopedia of
  social network analysis and mining (esnam), 2012.

\bibitem{gupta20131}
Aditi Gupta, Hemank Lamba, and Ponnurangam Kumaraguru.
\newblock \$1.00 per rt \#bostonmarathon \#prayforboston: Analyzing fake
  content on twitter.
\newblock In {\em eCrime Researchers Summit (eCRS), 2013}, pages 1--12. IEEE,
  2013.

\bibitem{govtech}
WAYNE HANSON.
\newblock How social media is changing law enforcement, December 2011.

\bibitem{hare2013twitter}
Jonathon~S Hare, Sina Samangooei, David~P Dupplaw, and Paul~H Lewis.
\newblock Twitter's visual pulse.
\newblock In {\em Proceedings of the 3rd ACM conference on International
  conference on multimedia retrieval}, pages 297--298. ACM, 2013.

\bibitem{hoi2011sire}
Steven~CH Hoi and Pengcheng Wu.
\newblock Sire: a social image retrieval engine.
\newblock In {\em Proceedings of the 19th ACM international conference on
  Multimedia}, pages 817--818. ACM, 2011.

\bibitem{kharroub2015social}
Tamara Kharroub and Ozen Bas.
\newblock Social media and protests: An examination of twitter images of the
  2011 egyptian revolution.
\newblock {\em New Media \& Society}, page 1461444815571914, 2015.

\bibitem{mcparlane2014picture}
Philip~J McParlane, Andrew~James McMinn, and Joemon~M Jose.
\newblock Picture the scene...;: Visually summarising social media events.
\newblock In {\em Proceedings of the 23rd ACM International Conference on
  Conference on Information and Knowledge Management}, pages 1459--1468. ACM,
  2014.

\bibitem{mendoza2010twitter}
Marcelo Mendoza, Barbara Poblete, and Carlos Castillo.
\newblock Twitter under crisis: can we trust what we rt?
\newblock In {\em Proceedings of the first workshop on social media analytics},
  pages 71--79. ACM, 2010.

\bibitem{kul}
BBC News.
\newblock Mumbai ink attack: India police arrest six shiv sena workers, October
  2015.

\bibitem{pass1997comparing}
Greg Pass, Ramin Zabih, and Justin Miller.
\newblock Comparing images using color coherence vectors.
\newblock In {\em Proceedings of the fourth ACM international conference on
  Multimedia}, pages 65--73. ACM, 1997.

\bibitem{popescu2011social}
Adrian Popescu and Gregory Grefenstette.
\newblock Social media driven image retrieval.
\newblock In {\em Proceedings of the 1st ACM International Conference on
  Multimedia Retrieval}, page~33. ACM, 2011.

\bibitem{radford2015unsupervised}
Alec Radford, Luke Metz, and Soumith Chintala.
\newblock Unsupervised representation learning with deep convolutional
  generative adversarial networks.
\newblock {\em arXiv preprint arXiv:1511.06434}, 2015.

\bibitem{Rogers_2014}
Simon Rogers.
\newblock What fuels a tweet's engagement, March 2014.

\bibitem{rublee2011orb}
Ethan Rublee, Vincent Rabaud, Kurt Konolige, and Gary Bradski.
\newblock Orb: An efficient alternative to sift or surf.
\newblock In {\em 2011 International conference on computer vision}, pages
  2564--2571. IEEE, 2011.

\bibitem{ruddell2013social}
Rick Ruddell and Nicholas Jones.
\newblock Social media and policing: matching the message to the audience.
\newblock {\em Safer Communities}, 12(2):64--70, 2013.

\bibitem{sachdeva2015online}
Niharika Sachdeva and Ponnurangam Kumaraguru.
\newblock Online social networks and police in india -understanding the
  perceptions, behavior, challenges.
\newblock In {\em ECSCW 2015: Proceedings of the 14th European Conference on
  Computer Supported Cooperative Work, 19-23 September 2015, Oslo, Norway},
  pages 183--203. Springer, 2015.

\bibitem{sachdeva2015social}
Niharika Sachdeva and Ponnurangam Kumaraguru.
\newblock Social networks for police and residents in india: exploring online
  communication for crime prevention.
\newblock In {\em Proceedings of the 16th Annual International Conference on
  Digital Government Research}, pages 256--265. ACM, 2015.

\bibitem{sachdeva2016social}
Niharika Sachdeva and Ponnurangam Kumaraguru.
\newblock Social media-new face of collaborative policing?
\newblock In {\em International Conference on Social Computing and Social
  Media}, pages 221--233. Springer, 2016.

\bibitem{solutions2012survey}
Lexis Nexis~Risk Solutions.
\newblock Survey of law enforcement personnel and their use of social media in
  investigations.
\newblock {\em Lexis Nexis}, 2012.

\bibitem{modi}
Business Standard.
\newblock Bjp files complaint against journalist for morphed modi photo, April
  2016.

\bibitem{starbird2011voluntweeters}
Kate Starbird and Leysia Palen.
\newblock Voluntweeters: Self-organizing by digital volunteers in times of
  crisis.
\newblock In {\em Proceedings of the SIGCHI Conference on Human Factors in
  Computing Systems}, pages 1071--1080. ACM, 2011.

\bibitem{shivaji}
Sakal Times.
\newblock 180 held for circulating obscene photos on internet, June 2014.

\bibitem{ecb_2014}
The~Economics Times.
\newblock Ncrb to connect police stations and crime data across country in 6
  months, November 2014.

\bibitem{kejri}
India Today.
\newblock Arvind kejriwal slammed on twitter again! twitterati trend
  \#kejriwalinsultshanuman, February 2016.

\bibitem{RR}
India Today.
\newblock Case against gurmeet ram rahim for posing as vishnu, will he be
  arrested?, January 2016.

\bibitem{tola2008fast}
Engin Tola, Vincent Lepetit, and Pascal Fua.
\newblock A fast local descriptor for dense matching.
\newblock In {\em Computer Vision and Pattern Recognition, 2008. CVPR 2008.
  IEEE Conference on}, pages 1--8. IEEE, 2008.

\bibitem{tola2010daisy}
Engin Tola, Vincent Lepetit, and Pascal Fua.
\newblock Daisy: An efficient dense descriptor applied to wide-baseline stereo.
\newblock {\em IEEE transactions on pattern analysis and machine intelligence},
  32(5):815--830, 2010.

\bibitem{vieweg2010microblogging}
Sarah Vieweg, Amanda~L Hughes, Kate Starbird, and Leysia Palen.
\newblock Microblogging during two natural hazards events: what twitter may
  contribute to situational awareness.
\newblock In {\em Proceedings of the SIGCHI conference on human factors in
  computing systems}, pages 1079--1088. ACM, 2010.

\bibitem{wang2012automatic}
Xiaofeng Wang, Matthew~S Gerber, and Donald~E Brown.
\newblock Automatic crime prediction using events extracted from twitter posts.
\newblock In {\em International Conference on Social Computing,
  Behavioral-Cultural Modeling, and Prediction}, pages 231--238. Springer,
  2012.

\bibitem{wiegand2016veracity}
Stefanie Wiegand and Stuart~E Middleton.
\newblock Veracity and velocity of social media content during breaking news:
  Analysis of november 2015 paris shootings.
\newblock In {\em Proceedings of the 25th International Conference Companion on
  World Wide Web}, pages 751--756. International World Wide Web Conferences
  Steering Committee, 2016.

\bibitem{yu2015improved}
Lei Yu, Zhixin Yu, and Yan Gong.
\newblock An improved orb algorithm of extracting and matching.
\newblock 2015.

\end{thebibliography}

\end{document}